%% file: Klaus-2.tex
\documentclass[twoside]{elsart}
\usepackage{epsfig}
\usepackage{array}
\usepackage{amssymb,amsbsy,amsmath}
\input{def.tex}
\begin{document}
%---------- Journal-----------------------------------------
%
\journal{PHYSICA A}
%
%---------- Titel und Abstract-------------------------------
%
\begin{frontmatter}
\title{Quasi-Particle Picture for Monatomic Gases}
\author{K. B\"arwinkel, J. Schnack\thanksref{JS}}
\author{, U. Thelker\thanksref{UT}}
\address{Universit\"at Osnabr\"uck, Fachbereich Physik \\ 
         Barbarastr. 7, D-49069 Osnabr\"uck}
\thanks[JS]{Email: jschnack\char'100uos.de,
            WWW:~http://obelix.physik.uni-osnabrueck.de/$\sim$schnack}
\thanks[UT]{present address: Westfalenring 58, D-49565 Bramsche}
\maketitle

\begin{abstract}
\noindent
A quasi-particle theory for monatomic gases in equilibrium is
formulated and evaluated to yield the exact virial contributions
to the thermodynamic state functions in lowest order of the
density. Van der Waals blocking has necessarily to be accounted
for in occupation number statistics. The quasi-particle
distribution function differs from the Wigner function by a
bilinear functional thereof. The progress made so far is
promising with respect to a corresponding version of kinetic
theory. 
\\
\vspace{1ex}
\\
\vspace{1ex}
\end{abstract}
\end{frontmatter}
%%%%%%%%%%%%%%%%%%%%%%%%%%%%%%%%%%%%%%%%%%%%%%%%%%%%%%%%%%%%%%%%%%%%%%%%
%\newpage
%\tableofcontents
%%%%%%%%%%%%%%%%%%%%%%%%%%%%%%%%%%%%%%%%%%%%%%%%%%%%%%%%%%%%%%%%%%%%%%%%
%\newpage
%\raggedbottom
\input{Klaus-2-0.tex}
\input{Klaus-2-1.tex}

\input{Klaus-2-2.tex}
\input{Klaus-2-3.tex}
\input{Klaus-2-4.tex}
\input{Klaus-2-6.tex}

\input{Klaus-2-5.tex}

%\clearpage
%%%%%%%%%%%%%%%%%%%%%%%%%%%%%%%%%%%%%%%%%%%%%%%%%%%%%%%%%%%%%%%%%%%%%%%%
%{\bf Acknowledgments}\\[5mm]
%%%%%%%%%%%%%%%%%%%%%%%%%%%%%%%%%%%%%%%%%%%%%%%%%%%%%%%%%%%%%%%%%%%%%%%%

\input{Klaus-2-Bib.tex}
\end{document}

%% file: def.tex
\newcommand{\skipit}[1]{}
\newcommand{\xref}[1]{\protect\ref{#1}}

\newcommand{\fmref}[1]{(\protect\ref{#1})}
\newlength{\CaptionWidth}
\newcommand{\mycaption}[2]{%
\begin{center}\begin{minipage}{\CaptionWidth}%
\caption[]{#1}\label{#2}%
\end{minipage}\end{center}%
}%
\newcommand{\element}[2]{$^{#1}$#2}

\def\V0{\stackrel{\circ}{V}}
\def\v0{\stackrel{\circ}{v}}
\def\half{{\frac{1}{2}}\;}

\newcommand{\dint}{\mbox{d}}

\renewcommand{\Re}[1]{\mbox{Re}\left(#1\right)\;}
\renewcommand{\Im}[1]{\mbox{Im}\left(#1\right)\;}

\newcommand{\SmallMean}[1]{\langle\langle \; {#1}\; 
            \rangle\rangle}

\def\bra#1{\langle \, {#1} \, | \;}
\def\ket#1{\; | \, {#1} \, \rangle}
\newcommand{\braket}[2]{\langle \, {#1} \, | \, {#2} \, \rangle}

\def\erw#1{\,\langle \, {#1} \, \rangle\,}

\newcommand{\pp}[2]{\frac{\partial \, {#1}}{\partial \, {#2}}\;}
\newcommand{\dd}[2]{\frac{{d}\, {#1}}{{d} {#2}}\;}

\newcommand{\vek}[1]{{\!\vec{\,#1}}}

%% file: Klaus-2-0.tex
\section{Introduction and summary}

The quasi-particle concept serves as a camouflage of interaction
in many-body systems. The aim is to explain a macroscopic system
in equilibrium as an ensemble of noninteracting and countable
units called "quasi-particles". Non-equilibrium may then be
described by a kinetic theory for these quasi-particles.

In this article the interaction-free theory is occupation number
statistics which yields the entropy as a functional of the
quasi-particle distribution function $f$. Interaction between
gas atoms is taken care of by modelling the elementary cell
volume and by suitable constraints for $f$.

Our theory reproduces the exact quantum mechanical density
corrections to the distribution function and to equilibrium
thermodynamics.  The only partial success of related attempts
(e.g. \cite{RaS76}) stems from an insufficient handling of the
combinatorial entropy, i.e. the neglect of van der Waals
blocking.

%% file: Klaus-2-1.tex
\section{One-particle distribution function}

The one-particle distribution function $f(\vek{p}_1,\vek{r},t)$
of gas kinetics enables the calculation of densities (in
position space) for macroscopic quantities by taking moments
over momentum space. We insist the two simplest ones be
the number density $n$
%--------------------------------------------------------
\begin{eqnarray}
\label{E-1-1}
n(\vek{r},t) = \int \dint^3 p_1\; f(\vek{p}_1,\vek{r},t)
\end{eqnarray}
%--------------------------------------------------------
and the density of linear momentum $\vek{\pi}$
%--------------------------------------------------------
\begin{eqnarray}
\label{E-1-2}
\vek{\pi}(\vek{r},t) = \int \dint^3 p_1\; 
\vek{p}_1\; f(\vek{p}_1,\vek{r},t)
\ .
\end{eqnarray}
%--------------------------------------------------------
Any justification of gas kinetics within the frame of a more
fundamental theory starts with a definition of $f$ which one is
free to choose, provided the interpretation of the moments
\fmref{E-1-1} and \fmref{E-1-2} is correct. Our quantum
statistical ansatz reads
%--------------------------------------------------------
\begin{eqnarray}
\label{E-1-3}
f(\vek{p}_1,\vek{r},t)
:=
f_W(\vek{p}_1,\vek{r},t)
+
\Psi(\vek{p}_1,\vek{r},t)
\ ,
\end{eqnarray}
%--------------------------------------------------------
where $f_W$ is the Wigner distribution function 
%--------------------------------------------------------
\begin{eqnarray}
\label{E-1-4}
f_W(\vek{p}_1,\vek{r},t)
=
\left( 2 \pi \hbar\right)^{-3}
\int \dint^3 r^\prime
\erw{\psi^\dagger(\vek{r}+\frac{\vek{r}^\prime}{2},t)
     \psi(\vek{r}-\frac{\vek{r}^\prime}{2},t)}
\exp\left\{\frac{i}{\hbar}\;\vek{p}_1\cdot\vek{r}^\prime\right\}
\end{eqnarray}
%--------------------------------------------------------
and $\Psi$ is a bilinear functional thereof
%--------------------------------------------------------
\begin{eqnarray}
\label{E-1-5}
\hspace*{-10mm}&&\Psi(\vek{p}_1,\vek{r},t) :=\\
\hspace*{-10mm}&&
\int \dint^3 p_2 \dint^3 p_1^\prime \dint^3 p_2^\prime\;
E(\vek{p} , \vek{p}^\prime)
\delta\left(\vek{P}-\vek{P}^\prime\right)
\left[
f_W(\vek{p}_1,\vek{r},t) f_W(\vek{p}_2,\vek{r},t)
-
f_W(\vek{p}_1^\prime,\vek{r},t) f_W(\vek{p}_2^\prime,\vek{r},t)
\right]
\nonumber \\
\hspace*{-10mm}&&
+ 
\int \dint^3 p_2 \dint^3 p_1^\prime \dint^3 p_2^\prime\;
O(\vek{p} , \vek{p}^\prime)
\delta\left(\vek{P}-\vek{P}^\prime\right)
\left[
f_W(\vek{p}_1,\vek{r},t) f_W(\vek{p}_2,\vek{r},t)
+
f_W(\vek{p}_1^\prime,\vek{r},t) f_W(\vek{p}_2^\prime,\vek{r},t)
\right]
\nonumber
\ ,
\end{eqnarray}
%--------------------------------------------------------
with $\vek{P}, \vek{P}^\prime$ and $\vek{p}, \vek{p}^\prime$
denoting centre-of-mass and relative momenta throughout the
article 
%--------------------------------------------------------
\begin{eqnarray}
\label{E-1-10}
\vek{P} = \vek{p}_1 + \vek{p}_2 
\ , \qquad
\vek{p} = \half(\vek{p}_1 - \vek{p}_2 )
\ .
\end{eqnarray}
%--------------------------------------------------------
The even kernel 
%--------------------------------------------------------
\begin{eqnarray}
\label{E-1-6}
E(\vek{p} , \vek{p}^\prime)
=
E(\vek{p}^\prime , \vek{p})
=
E(-\vek{p} , -\vek{p}^\prime)
\end{eqnarray}
%--------------------------------------------------------
and the odd kernel 
%--------------------------------------------------------
\begin{eqnarray}
\label{E-1-7}
O(\vek{p} , \vek{p}^\prime)
=
-O(\vek{p}^\prime , \vek{p})
=
O(-\vek{p} , -\vek{p}^\prime)
\end{eqnarray}
%--------------------------------------------------------
remain to be specified. The structure of $\Psi$ reminds of
Boltzmann's collision integral although there is no
$\delta$-function for the kinetic energies of relative motion
$E_p$, $E_{p^\prime}$ which are defined by
%--------------------------------------------------------
\begin{eqnarray}
\label{E-1-11}
E_p = \frac{\vek{p}^2}{2 m_{rel}} = \frac{\vek{p}^2}{m}
\ .
\end{eqnarray}
%--------------------------------------------------------
In any case equations \fmref{E-1-1} and \fmref{E-1-2} hold:
%--------------------------------------------------------
\begin{eqnarray}
\label{E-1-8}
n(\vek{r},t)
=
\erw{\psi^\dagger(\vek{r},t)\psi(\vek{r},t)}
=
\int \dint^3 p_1\; f(\vek{p}_1,\vek{r},t)
=
\int \dint^3 p_1\; f_W(\vek{p}_1,\vek{r},t)
\end{eqnarray}
%--------------------------------------------------------
and
%--------------------------------------------------------
\begin{eqnarray}
\label{E-1-9}
\vek{\pi}(\vek{r},t)
=
\int \dint^3 p_1\; \vek{p}_1 \; f(\vek{p}_1,\vek{r},t)
=
\int \dint^3 p_1\; \vek{p}_1 \; f_W(\vek{p}_1,\vek{r},t)
\ .
\end{eqnarray}
%--------------------------------------------------------
There is not only a quantum statistical motivation to the
definition \fmref{E-1-3}, but it also allows a quasi-particle
interpretation.

%% file: Klaus-2-2.tex
\section{Quantum statistical background}

Explicit expressions for the kernels $E$ and $O$ follow from the
theory of Kadanoff and Baym \cite{KaB62}, from which in the spatially
homogeneous case and in ${\mathcal T}$-matrix approximation the
kinetic equation
%--------------------------------------------------------
\begin{eqnarray}
\label{E-2-1}
\partial_t (f_W + \Psi) = J_{B}[f_W]
\end{eqnarray}
%--------------------------------------------------------
is obtained \cite{Bae69,Bae84} with $J_{B}[f_W]$ as Boltzmann's
collision integral depending on $f_W$. Here $\Psi$ is just
the functional \fmref{E-1-5} with the kernels
%--------------------------------------------------------
\begin{eqnarray}
\label{E-2-2}
O(\vek{p} , \vek{p}^\prime)
&=&
\frac{1}{4}(2\pi\hbar)^3
{\mathcal P}(E_p - E_{p^\prime})
\\
&&
\left\{
|\bra{\vek{p}}{\mathcal T}_{\pm}(E_{{p}^\prime}+i\epsilon)
 \ket{\vek{p}^\prime}|^2
-
|\bra{\vek{p}^\prime}{\mathcal T}_{\pm}(E_{{p}}+i\epsilon)
 \ket{\vek{p}}|^2
\right\}
\nonumber
\end{eqnarray}
%--------------------------------------------------------
and
%--------------------------------------------------------
\begin{eqnarray}
\label{E-2-3}
E(\vek{p} , \vek{p}^\prime)
=&&
\pi (2\pi\hbar)^3
\delta(E_p - E_{p^\prime})
\\
&&
\Im{
   \bra{\vek{p}}{\mathcal T}_{\pm}(E_{{p}^\prime}+i\epsilon)
   \ket{\vek{p}^\prime}^*
   \bra{\vek{p}}{\mathcal T}_{\pm}^\prime(E_{{p}^\prime}+i\epsilon)
   \ket{\vek{p}^\prime}
   }
\nonumber \\
&+&
\frac{1}{4}(2\pi\hbar)^3
{\mathcal P}^\prime(E_p - E_{p^\prime})
\nonumber
\\
&&
\left\{
|\bra{\vek{p}}{\mathcal T}_{\pm}(E_{p^\prime}+i\epsilon)
 \ket{\vek{p}^\prime}|^2
+
|\bra{\vek{p}^\prime}{\mathcal T}_{\pm}(E_{p}+i\epsilon)
 \ket{\vek{p}}|^2
\right\}
\nonumber
\ .
\end{eqnarray}
%--------------------------------------------------------
The ${\mathcal T}$-matrix occurring here is the properly
symmetrized momentum representation of the two-particle operator
%--------------------------------------------------------
\begin{eqnarray}
\label{E-2-4}
{\mathcal T}(z)
=
V - V \frac{1}{H-z} V
\ , \qquad
{\mathcal T}^\prime(z) = \dd{}{z} {\mathcal T}(z)
\ ,
\end{eqnarray}
%--------------------------------------------------------
with $H=H_{kin}+V$ being the Hamiltonian of relative motion.
${\mathcal P}$ is the principle value distribution and
${\mathcal P}^\prime$ it's derivative. The upper sign in
${\mathcal T}_{\pm}$ (and elsewhere) refers to bosons and the
lower sign to fermions.

In equilibrium, quantum statistical mechanics yields the density
expansion for $f=f_W+\Psi$, which up to second order reads
%--------------------------------------------------------
\begin{eqnarray}
\label{E-2-6}
f_{eq}(\vek{p}_1)
=
\frac{n}{(2\pi m \kappa T)^{3/2}}
e^{-\frac{\vek{p}_1^2}{2 m \kappa T}}
\left(
1 + n (2 B(T) + \phi(\vek{p}_1))
\right)
\ ,
\end{eqnarray}
%--------------------------------------------------------
with $\kappa$ being Boltzmann's constant and 
%--------------------------------------------------------
\begin{eqnarray}
\label{E-2-7}
\phi(\vek{p}_1)
&=&
\pm 
\lambda^3(T)\; e^{-\frac{\vek{p}_1^2}{2 m \kappa T}}
-
\lambda^3(T)\; \int \dint^3 p_2\; 
e^{-\frac{\vek{p}_2^2}{2 m \kappa T}}
\Big\{
   \frac{\tilde{F}({p})}{\kappa T}
   +\tilde{G}({p})
\Big\}
\nonumber
\ .
\end{eqnarray}
%--------------------------------------------------------
Here we have introduced the thermal wave length
%--------------------------------------------------------
\begin{eqnarray}
\label{E-2-8}
\lambda
=
\frac{2\pi\hbar}{\sqrt{2\pi m \kappa T}}
\ .
\end{eqnarray}
%--------------------------------------------------------
The quantities $\tilde{F}$ and $\tilde{G}$ as well as the second
virial coefficient $B(T)$ are given as
functionals of the ${\mathcal T}$-matrix:
%---------------------------------------------------------------------
\begin{eqnarray}
\label{E-2-8-1}
\tilde{F}({p})
=
\half (2\pi\hbar)^3\;
\Re{\bra{\vek{p}}{\mathcal T}_{\pm}(E_{{p}}+i\epsilon)
\ket{\vek{p}}}
\ ,
\end{eqnarray}
%---------------------------------------------------------------------
%---------------------------------------------------------------------
\begin{eqnarray}
\label{E-2-8-2}
\tilde{G}({p})
=
\frac{\pi}{2} (2\pi\hbar)^3\;
\int \dint^3 q\; 
&&
\delta\left(E_{{p}} - E_{{q}} \right)
\\
&\times&
\Im{
\bra{\vek{p}}{\mathcal T}_{\pm}(E_{{q}}+i\epsilon)
\ket{\vek{q}}
\bra{\vek{q}}{\mathcal T}_{\pm}^\prime(E_{{q}}+i\epsilon)
\ket{\vek{p}}^*
}
\ ,
\nonumber
\end{eqnarray}
%---------------------------------------------------------------------
%--------------------------------------------------------
\begin{eqnarray}
\label{E-2-9}
B(T) &=& B_0(T) + B_1(T) + B_2(T)
\\[2mm]
B_0(T) &=& \mp 2^{-5/2}\; \lambda^3
\ , \quad
B_1(T) = \frac{\SmallMean{\tilde{F}}}{\kappa T}
\ , \quad
B_2(T) = \SmallMean{\tilde{G}}
\nonumber 
\ ,
\end{eqnarray}
%--------------------------------------------------------
where $\SmallMean{\cdot}$ denotes the thermal average, e.g.
%---------------------------------------------------------------------
\begin{eqnarray}
\label{E-2-9-1}
\SmallMean{\tilde{F}}
&=&
\frac{\int\dint^3 p\; \exp\{-\frac{E_p}{\kappa T}\} \tilde{F}(p)}
     {\int\dint^3 p\; \exp\{-\frac{E_p}{\kappa T}\} }
\ .
\end{eqnarray}
%---------------------------------------------------------------------
These formulae are exact if the two-particle interaction does
not allow any bound states. $B_1$ essentially accounts for
long-range attraction and $B_2$ for hard repulsion, this
correspondence being most striking in the van der Waals limit
(cf. section \xref{SecvdW}).

%% file: Klaus-2-3.tex
\section{Quasi-particle picture for equilibrium}

We just work out the usual idea:
\begin{enumerate}
\item The entropy density $s$ is represented as a functional of the
one-particle distribution function. This is an outcome of
occupation number statistics (combinatorial entropy)
%---------------------------------------------------------------------
\begin{eqnarray}
\label{E-3-1}
s = -\kappa \int \dint^3 p_1 \;
\Big[
&&
f(\vek{p}_1) \ln\big(v_{el}(\vek{p}_1)f(\vek{p}_1)\big)
\\
&&\mp
\left(
\frac{1}{v_{el}(\vek{p}_1)}\pm f(\vek{p}_1)
\right)
\ln\big(1\pm v_{el}(\vek{p}_1)f(\vek{p}_1)\big)
\Big]
\nonumber
\end{eqnarray}
%---------------------------------------------------------------------
with $v_{el}$ as the volume of an elementary cell in
six-dimensional $\mu$-space, $v_{el}$ accommodating one
single-particle quantum state. Eq. \fmref{E-3-1} is well known as
a standard result for non-interacting particles. However, the
choice of $v_{el}$ and the constraints for $f$ may provide a
camouflage of interaction.
\item For equilibrium the distribution function $f$ is the one
which minimizes $s$ subject to appropriate constraints.
\item The constraints are a given number density $n$,
%---------------------------------------------------------------------
\begin{eqnarray}
\label{E-3-2}
n = \int \dint^3 p_1\; f(\vek{p}_1)
\ ,
\end{eqnarray}
%---------------------------------------------------------------------
and a given energy density $u$,
%---------------------------------------------------------------------
\begin{eqnarray}
\label{E-3-3}
u = \int \dint^3 p_1\; \varepsilon(\vek{p}_1)\; f(\vek{p}_1)
\ .
\end{eqnarray}
%---------------------------------------------------------------------
\end{enumerate}

The quasi-particle interpretation is now introduced by way of
ansatz (eqs. \fmref{E-3-4}, \fmref{E-3-5}), its aim being to
account for interaction effects in lowest order of the
density. Strong repulsion reduces the freely accessible volume
for gas particles.  Because of this effect, called "van der
Waals blocking", more than just $(2\pi\hbar)^3$ is needed as an
elementary cell volume
%---------------------------------------------------------------------
\begin{eqnarray}
\label{E-3-4}
v_{el}(\vek{p}_1) = 
(2\pi\hbar)^3
\left[1 +
\int \dint^3 p_2\; 
G\left(
 \left|
 \vek{p}
 \right|
 \right)
\; f(\vek{p}_2)
\right]
\ .
\end{eqnarray}
%---------------------------------------------------------------------
Also, an interacting gas particle carries with it a correlation
cloud giving rise to an interactive contribution which changes
the kinetic energy of a particle into the energy of a quasi
particle
%---------------------------------------------------------------------
\begin{eqnarray}
\label{E-3-5}
\varepsilon(\vek{p}_1) = 
\frac{\vek{p}_1^2}{2 m}
+
\int \dint^3 p_2\; 
F\left(
 \left|
 \vek{p}
 \right|
 \right)
\; f(\vek{p}_2)
\ .
\end{eqnarray}
%---------------------------------------------------------------------
For our variational problem the functions $G$ and $F$ are
considered as given though, for the time being, unknown. They
will be determined afterwards by comparing the equilibrium
solution $f=f_{eq}$ with the corresponding expression from
many-body quantum theory. The resulting thermodynamics then
serves as a further touchstone of the quasi-particle
interpretation. 

The solution of the variational problem is obviously equivalent
to 
%---------------------------------------------------------------------
\begin{eqnarray}
\label{E-3-6}
\left(
\frac{\delta s}{\delta f(\vek{p})}
\right)_{f=f_{eq}}
=
\frac{1}{\vartheta}
\left(
\frac{\delta u}{\delta f(\vek{p})}
-
\alpha
\frac{\delta n}{\delta f(\vek{p})}
\right)_{f=f_{eq}}
\end{eqnarray}
%---------------------------------------------------------------------
with $\vartheta$ and $\alpha$ as Lagrange parameters due to the the
constraints for $f$. Comparison with the thermodynamic identity
for the entropy $S$ at constant volume 
%---------------------------------------------------------------------
\begin{eqnarray}
\label{E-3-7}
\dint S
=
\frac{1}{T}
\left(
\dint U - \mu \dint N
\right)
\end{eqnarray}
%---------------------------------------------------------------------
reveals that $\vartheta$ means the temperature, $\vartheta=T$,
and $\alpha$ means the chemical potential,
$\alpha=\mu$. According to \fmref{E-3-6} $f=f_{eq}$ is
equivalently determined by the fixed-point equation
%---------------------------------------------------------------------
\begin{eqnarray}
\label{E-3-8}
f(\vek{p}_1) =
\frac{\zeta}{v_{el}(\vek{p}_1)}\;
\frac{\exp\left\{-\beta\left(\frac{\vek{p}_1^2}{2 m}+K(\vek{p}_1) 
          \right)\right\}}
     {1\mp \zeta \exp\left\{-\beta\left(\frac{\vek{p}_1^2}{2 m}+K(\vek{p}_1) 
                   \right)\right\}}
\end{eqnarray}
%---------------------------------------------------------------------
with
%---------------------------------------------------------------------
\begin{eqnarray}
\label{E-3-9}
\beta = \frac{1}{\kappa T}
\ , \qquad
\zeta = \exp\left\{\beta\mu\right\}
\end{eqnarray}
%---------------------------------------------------------------------
and
%---------------------------------------------------------------------
\begin{eqnarray}
\label{E-3-10}
K(\vek{p}_1) = 
\int \dint^3 p_2\;
&\Bigg\{&
2 f(\vek{p}_2)
F\left(
 \left|
 \vek{p}
 \right|
 \right)
\\
&&
\pm
\frac{(2\pi\hbar)^3}{\beta v_{el}^2(\vek{p}_2)}\;
\ln\big(1\pm v_{el}(\vek{p}_2)f(\vek{p}_2)\big)
G\left(
 \left|
 \vek{p}
 \right|
 \right)
\Bigg\}
\nonumber
\ .
\end{eqnarray}
%---------------------------------------------------------------------

%% file: Klaus-2-4.tex
\section{Lowest-order density corrections}

The fugacity $\zeta$ can be given as a power series in $n$
%---------------------------------------------------------------------
\begin{eqnarray}
\label{E-4-1}
\zeta = n \lambda^3
\left(
1 + 2 n B(T) + \cdots
\right)
\end{eqnarray}
%---------------------------------------------------------------------
and the fixed point equation \fmref{E-3-8} may be iterated
starting off with a Maxwellian normalized to $n$. This yields a
density expansion according to which -- apart from third and
higher order contributions -- one regains eq. \fmref{E-2-6} for
$f$, but now with
%--------------------------------------------------------
\begin{eqnarray}
\label{E-4-2}
\phi(\vek{p}_1)
&=&
\pm 
\lambda^3\; e^{-\frac{\beta}{2} E_{{p}_1}}
\\
&&
-
\frac{2 \lambda^3}{(2\pi\hbar)^3}
\; \int \dint^3 p_2\; 
e^{-\frac{\beta}{2} E_{{p}_2}}
\left\{
\beta
F\left(
 \left|
 \vek{p}
 \right|
 \right)
+
G\left(
 \left|
 \vek{p}
 \right|
 \right)
\right\}
\ .
\nonumber
\end{eqnarray}
%--------------------------------------------------------
Therefore, the quasi-particle picture independently explains
eq. \fmref{E-2-6}, if and only if
%---------------------------------------------------------------------
\begin{eqnarray}
\label{E-4-3}
F({p})
=
\tilde{F}({p})
\end{eqnarray}
%---------------------------------------------------------------------
and
%---------------------------------------------------------------------
\begin{eqnarray}
\label{E-4-4}
G({p})
=
\tilde{G}({p})
\end{eqnarray}
%---------------------------------------------------------------------
Having determined $f$ up to second order in $n$, we deduce the
entropy with its lowest order density corrections from
eq. \fmref{E-3-1}:
%---------------------------------------------------------------------
\begin{eqnarray}
\label{E-4-5}
s = n \kappa
\left\{
\frac{5}{2} - \ln\left( n \lambda^3 \right)
- n \left[ B(T) + T B^\prime(T) \right]
\right\}
\ ,
\end{eqnarray}
%---------------------------------------------------------------------
with $B(T)$ defined by eqs. \fmref{E-2-9}.
Analogously evaluating eq. \fmref{E-3-3}, we obtain the two
leading contributions to the energy density:
%---------------------------------------------------------------------
\begin{eqnarray}
\label{E-4-6}
u = n \kappa T
\left\{
\frac{3}{2} - n T B^\prime(T)
\right\}
\ .
\end{eqnarray}
%---------------------------------------------------------------------
The last two results imply the pressure equation of state
%---------------------------------------------------------------------
\begin{eqnarray}
\label{E-4-8}
p = n 
\left(
\pp{u}{n}
\right)_{s/n}
- u
=
n \kappa T
\left\{
1 + n B(T)
\right\}
\ .
\end{eqnarray}
%---------------------------------------------------------------------
The exact density corrections, i.e. virial contributions have
thus been obtained.

%% file: Klaus-2-6.tex
\section{Classical van der Waals approximation}
\label{SecvdW}

Considering distinguishable particles, one has to neglect
quantum statistical contributions. This means
%---------------------------------------------------------------------
\begin{eqnarray}
\label{E-5-1}
B_0 = 0
\ .
\end{eqnarray}
%---------------------------------------------------------------------
Then with \fmref{E-2-9}, \fmref{E-4-3} and \fmref{E-4-4}
%---------------------------------------------------------------------
\begin{eqnarray}
\label{E-5-2}
B(T)
&=&
\frac{\SmallMean{F}}{\kappa T}
+ \SmallMean{G}
\end{eqnarray}
%---------------------------------------------------------------------
holds and suggests a comparison with the van der
Waals version of the second virial coefficient
%---------------------------------------------------------------------
\begin{eqnarray}
\label{E-5-4}
B_{vdW}(T)
&=&
- \frac{a}{\kappa T}
+ b
\ ,
\end{eqnarray}
%---------------------------------------------------------------------
which is readily obtained from the model equation of state
%---------------------------------------------------------------------
\begin{eqnarray}
\label{E-5-5}
\left(p + n^2 a \right)
\left(1 - n b \right)
=
n \kappa T
\end{eqnarray}
%---------------------------------------------------------------------
and the corresponding density expansion
%---------------------------------------------------------------------
\begin{eqnarray}
\label{E-5-6}
p
&=&
n \kappa T \left(1 + n \left[ b - \frac{a}{\kappa T}\right]
\right) + o(n^3)
\ .
\end{eqnarray}
%---------------------------------------------------------------------
The van der Waals limit therefore obviously means 
%---------------------------------------------------------------------
\begin{eqnarray}
\label{E-5-7}
\SmallMean{F}
&=&
\mbox{const}
=
- a
\qquad\mbox{and}
\qquad
\SmallMean{G}
=
\mbox{const}
=
b
\ .
\end{eqnarray}
%---------------------------------------------------------------------
This model assumption is actually quite reasonable as we are
going to demonstrate for \element{4}{He} atoms interacting via a
Lennard-Jones-potential lacking bound states \cite{The90}:
%---------------------------------------------------------------------
\begin{eqnarray}
\label{E-5-9}
V(r)
&=&
4 V_0
\left[
\left(
\frac{\sigma}{r}
\right)^{12}
-
\left(
\frac{\sigma}{r}
\right)^{6}
\right]
\ ;\quad
V_0 = 10.22\,\mbox{K} \; \kappa
\quad , \quad
\sigma = 2.56\,\mbox{\AA}
\ .
\end{eqnarray}
%---------------------------------------------------------------------
Because the interaction is radially symmetric the following
relations between the (anti-) symmetrized ${\mathcal T}$-matrix, the scattering
amplitude $f_{\pm}(p,\theta)$ and phase shifts $\delta_l(p)$
%---------------------------------------------------------------------
\begin{eqnarray}
\label{E-5-11}
f_{\pm}(p,\theta)
&=&
- \pi^2 m \hbar 
\bra{\vek{p}}{\mathcal T}_{\pm}(E_{{p}}+i\epsilon)\ket{\vek{q}}
\ , \
|\vek{p}|=|\vek{q}|\ ,\ \vek{p}\cdot\vek{q}=|\vek{p}|
|\vek{q}| \cos(\theta)
\\
f_{\pm}(p,\theta)
&=&
\frac{\hbar}{p}
\sum_l{}^\prime\; (2l+1)\; \mbox{e}^{i\, \delta_l(p)}\; \sin(\delta_l(p))\; P(\cos(\theta))
\end{eqnarray}
%---------------------------------------------------------------------
may be used (see e.g. \cite{Bau67}), where the summation runs
over even $l$ for bosons and odd $l$ for fermions. $F(p)$ and
$G(p)$ (compare to eqs. \fmref{E-2-8-1} and \fmref{E-2-8-2}) can
then be expressed in terms of phase shifts which is compatible
with the Beth Uhlenbeck result for $B(T)$ \cite{BeU36}
%---------------------------------------------------------------------
\begin{eqnarray}
\label{E-5-10}
F(p)
&=&
-\frac{4 \pi \hbar^2}{m} f_{\vek{p}}(0)
=
-\frac{4 \pi \hbar^2}{m} 
\frac{\hbar}{2 p}
\sum_l{}^\prime\; (2l + 1) \sin\left[ 2 \delta_l(p) \right]
\\
G(p)
&=&
-\hbar \int\dint\Omega\; \mbox{Im}
\left[ (f_{\vek{p}}(\theta))^* \pp{}{p} f_{\vek{p}}(\theta)
\right]
=
-{4 \pi \hbar}
\frac{\hbar^2}{p^2}
\sum_l{}^\prime\; (2l + 1) \sin^2\left[ \delta_l(p) \right]
\pp{\delta_l}{p}
\ .
\nonumber
\end{eqnarray}
%---------------------------------------------------------------------

%===================    figure   =================================
\begin{figure}[ht!]
\unitlength1mm
\begin{center}
\begin{picture}(150,55)
\put(-5,0){\epsfig{file=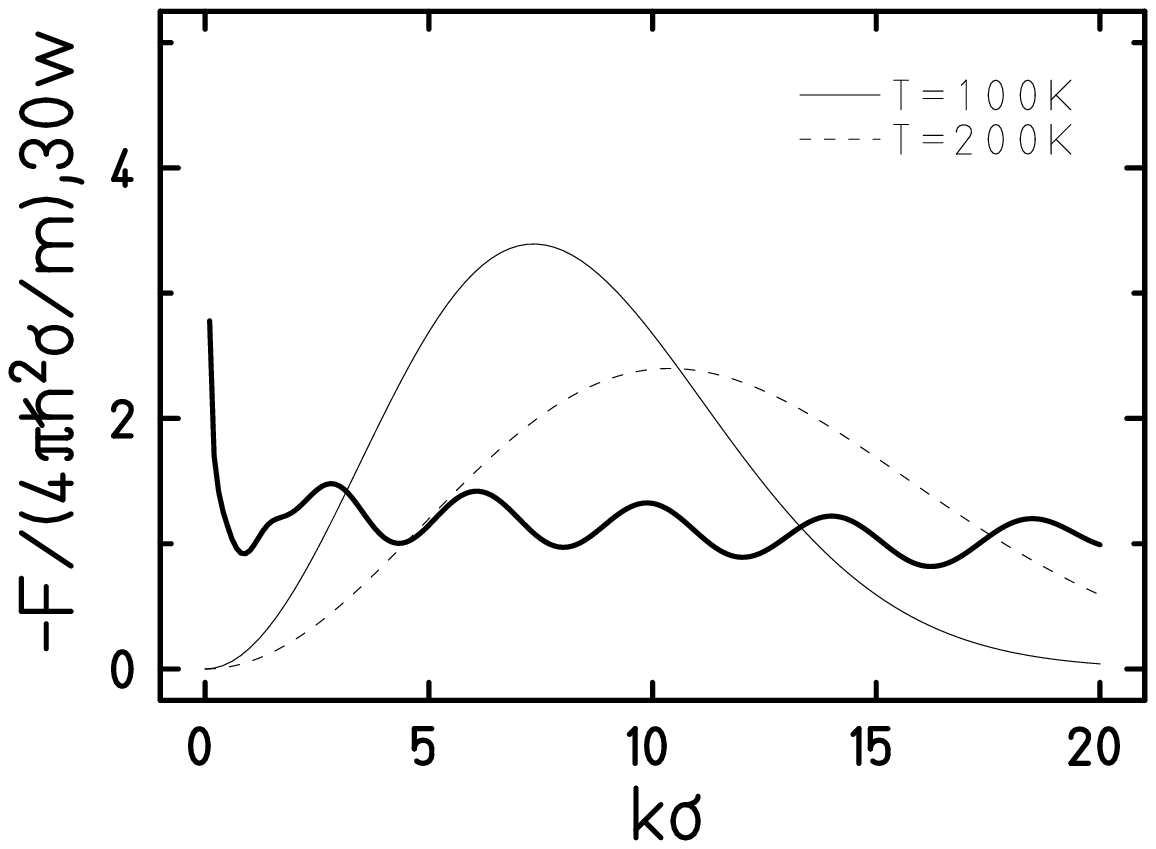,width=70mm}}
\put(75,0){\epsfig{file=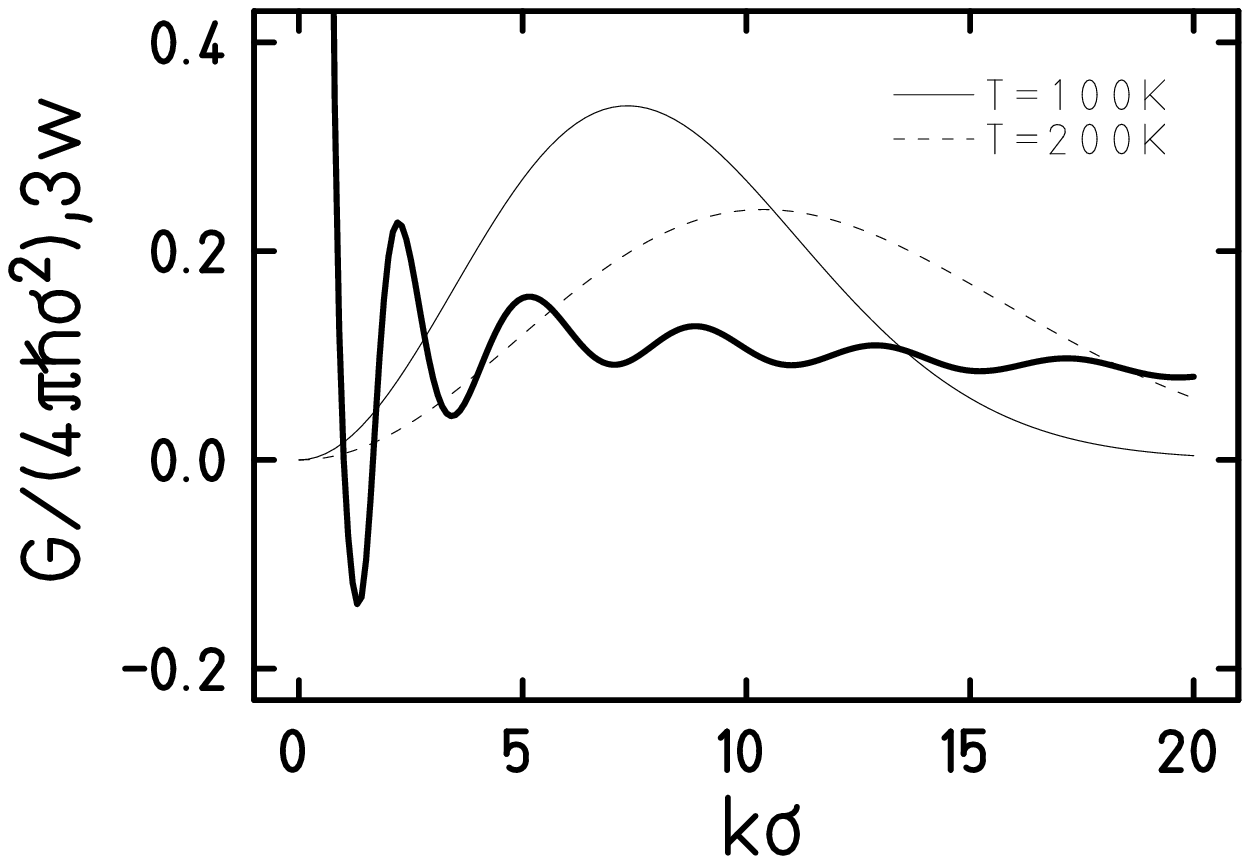,width=75mm}}
\end{picture}
\end{center}
\mycaption{$F$ and $G$ as functions of $k\sigma$ (thick solid
lines) and thermal weight functions 
$w\propto p^2 \exp\left\{ -E_p/(\kappa T)\right\}$ for two
temperatures.}{F-6-1}
\end{figure} 
%===================    figure   =================================

Figure \xref{F-6-1} nicely shows that for a large region of
temperatures $\SmallMean{F}$ and $\SmallMean{G}$ may be nearly
considered as constants, i.e. not depending on temperature.  In
this case the single particle energy, the elementary cell
volume and the mean energy are easily determined as
%---------------------------------------------------------------------
\begin{eqnarray}
\label{E-5-12}
\varepsilon(\vek{p}_1) = 
\frac{\vek{p}_1^2}{2 m}
- a n
\ , \quad
v_{el}(\vek{p}_1) = 
(2\pi\hbar)^3
\left[1 + b n \right]
\ , \quad
u = n \kappa T
\left\{
\frac{3}{2} - \frac{n a}{\kappa T}
\right\}
\ .
\end{eqnarray}
%---------------------------------------------------------------------

%% file: Klaus-2-5.tex
\section{Physical meaning of the quasi-particles}

One may imagine a quasi-particle to be a gas particle together
with its surrounding correlation cloud which is described by
the radial distribution function $g(r,T)$. This interpretation
suggests itself because the virial correction to the energy of
the ideal gas (eq. \fmref{E-4-6}) is mainly determined by $g(r,T)$ in
an obvious way.  The interpretation is immediately
evident in the classical case where 
%---------------------------------------------------------------------
\begin{eqnarray}
\label{E-5-A}
B(T) 
&=&
B_{cl}(T) 
=
-\half \int \dint^3 r \;(g_0(r,T)-1)
\end{eqnarray}
%---------------------------------------------------------------------
with
%---------------------------------------------------------------------
\begin{eqnarray}
\label{E-5-B}
g_0(r,T)
&=&
g_{0,cl}(r,T)
=
\exp\left\{-\frac{V(r)}{\kappa T} \right\}
\end{eqnarray}
%---------------------------------------------------------------------
as the first term of the density expansion
%---------------------------------------------------------------------
\begin{eqnarray}
\label{E-5-C}
g(r,T) 
&=&
g_0(r,T) + n g_1(r,T) + n^2 g_2(r,T) + \cdots
\ .
\end{eqnarray}
%---------------------------------------------------------------------
Therefore
%---------------------------------------------------------------------
\begin{eqnarray}
\label{E-5-D}
n^2 \kappa T
\;
T B_{cl}^\prime(T)
&=&
\frac{n^2}{2}\int\dint^3 r \; g_{0,cl}(r,T) V(r)
\end{eqnarray}
%---------------------------------------------------------------------
is the classical virial correction to the internal energy.

In general, however, for a homogeneous system
%---------------------------------------------------------------------
\begin{eqnarray}
\label{E-5-E}
g(r,T) 
&=&
\frac{1}{n^2}
\SmallMean{\psi^\dagger(\vek{r}^\prime)\,\psi^\dagger(\vek{r}^\prime+\vek{r})\,
\psi(\vek{r}^\prime+\vek{r})\,\psi(\vek{r}^\prime)}
\\
&=&
\frac{2}{n}
\frac{\delta f_{free}}{\delta V(r)}
\end{eqnarray}
%---------------------------------------------------------------------
with $f_{free}$ being the free energy per particle, whence
%---------------------------------------------------------------------
\begin{eqnarray}
\label{E-5-F}
g_0(r,T)
&=&
2 \kappa T \frac{\delta B(T)}{\delta V(r)}
\ .
\end{eqnarray}
%---------------------------------------------------------------------
Then quantum mechanically \cite{BaG70,Bae84}
%---------------------------------------------------------------------
\begin{eqnarray}
\label{E-5-G}
\int\dint^3 r \; g_{0}(r,T) V(r)
&=& 
2^{3/2} \lambda^3 \int\dint^3 p\;
e^{-\beta E_p}\; {}^{(-)}\bra{\vek{p}}V_{\pm}\ket{\vek{p}}^{(-)}
\ ,
\end{eqnarray}
%---------------------------------------------------------------------
where the scattering eigenstates $\ket{\vek{p}}^{(-)}$ satisfy the
Lippmann Schwinger equation which reads in momentum
representation
%---------------------------------------------------------------------
\begin{eqnarray}
\label{E-5-H}
\braket{\vek{p}^\prime}{\vek{p}}^{(-)}
&=&
\delta(\vek{p}^\prime - \vek{p})
- (E_{p^\prime}-E_p-i\epsilon)
\bra{\vek{p}^\prime}V\ket{\vek{p}}^{(-)}
\ .
\end{eqnarray}
%---------------------------------------------------------------------
The virial correction to the energy density (eqs. \fmref{E-3-3},
\fmref{E-4-6})is made up of three terms
%---------------------------------------------------------------------
\begin{eqnarray}
\label{E-5-I}
n^2 \kappa T
\;
T B^\prime(T)
&=&
\int\dint^3 p_1\;
\frac{\vek{p}_1^2}{2 m} \Psi_M(\vek{p}_1)
+ 
\int\dint^3 p_1\;\dint^3 p_2\;
f_M(\vek{p}_1)\;f_M(\vek{p}_2)\;F(p)
\\
&&
+
\int\dint^3 p_1\;
\frac{\vek{p}_1^2}{2 m} n^2 f_{W,2}(\vek{p}_1)
\nonumber
\ .
\end{eqnarray}
%---------------------------------------------------------------------
Here $f_M$ is the Maxwellian normalized to $n$ and $\Psi_M$ is
our functional \fmref{E-1-5} with $f_W$ replaced by
$f_M$. $f_{W,2}$ denotes the second term in the density
expansion of the Wigner function
%---------------------------------------------------------------------
\begin{eqnarray}
\label{E-5-J}
f_{W}(\vek{p}_1)
&=&
f_{M}(\vek{p}_1) + n^2  f_{W,2}(\vek{p}_1) + \cdots
\ .
\end{eqnarray}
%---------------------------------------------------------------------
Looking more closely at eq. \fmref{E-5-I} and taking account of
\fmref{E-5-G} one can show that  
%---------------------------------------------------------------------
\begin{eqnarray}
\label{E-5-K}
\int\dint^3 p_1\;
\frac{\vek{p}_1^2}{2 m} \Psi_M(\vek{p}_1)
+ 
\int\dint^3 p_1\;\dint^3 p_2\;
f_M(\vek{p}_1)\;f_M(\vek{p}_2)\;F(p)
&=&
\int\dint^3 r \; g_{0}(r,T) V(r)
\ .
\end{eqnarray}
%---------------------------------------------------------------------
Therefore, in the classical limit the last term in
eq. \fmref{E-5-I} must vanish.

%% file: Klaus-2.bbl
\begin{thebibliography}{99}
%------------------------------------------------
\bibitem{RaS76}
	J.C. Rainwater, R.F. Snider,
	Phys. Rev. {\bf A13} (1976) 1190
%
%	
%
%------------------------------------------------
\bibitem{KaB62}
	L.P. Kadanoff, G. Baym,
	Quantum Statistical Mechanics,
	W.A. Benjamin, New York (1962)
%
%	
%
%------------------------------------------------
\bibitem{Bae69}
	K. B\"arwinkel,
	Z. f. Naturforschung,	
	{\bf 28a} (1969) 22
%
%	
%
%------------------------------------------------
\bibitem{Bae84}
	K. B\"arwinkel,
	Rarefied Gas Dynamics (ed. H.~Oguchi),
	University of Tokyo Press
	(1984) 3
%
%	
%
%------------------------------------------------
\bibitem{The90}
	J.E. Kilpatrick, M.F. Kilpatrick,
	J. Chem. Phys. {\bf 19} 1951 930; \\
	U. Thelker,
	diploma thesis,
	University of Osnabr\"uck (1990)
%
%	
%
%------------------------------------------------
\bibitem{Bau67}
	B.J. Baumgartl,
	Z. Phys. {\bf 198} (1967) 148
%
%	2. & 3. virial coefficient
%
%------------------------------------------------
\bibitem{BeU36}
	G.E. Uhlenbeck, E. Beth, 
	Physica {\bf 3} (1936) 729;\\
	E. Beth, G.E. Uhlenbeck, 
	Physica {\bf 4} (1937) 915
%
%	
%
%------------------------------------------------
\bibitem{BaG70}
	K. B\"arwinkel, S. Gro{\ss}mann, 
	Z. Phys. {\bf 230} (1970) 141
%
%	pair distribution function g(r;n,T)
%
%------------------------------------------------
%\bibitem{}
%	{\bf }
%
%	
%
%------------------------------------------------
\end{thebibliography}
